\documentclass[25pt]{article}
\usepackage{amsfonts,amsmath,amssymb,amsthm,epsfig,mathrsfs,subfigure}
\usepackage{setspace}
\theoremstyle{definition}

\theoremstyle{plain}

\title{Riemann-Hilbert approach for multi-soliton solutions of a fifth-order nonlinear Schr\"{o}dinger equation\footnote{This work was supported by the National Natural Science Foundation of China (Grant Nos. 61072147 and 11271008).}}
\author{Zhou-Zheng Kang$^{1,2}$, Tie-Cheng Xia$^{1}$\footnote{Corresponding author. E-mail: xiatc@shu.edu.cn.}\ , Xi Ma$^{1}$ \\
1. Department of Mathematics, Shanghai University, Shanghai 200444, China;\\
2. College of Mathematics, Inner Mongolia University for Nationalities,\\ Tongliao 028043, China\\}
\date{}
\begin{document}
\sloppy \maketitle
\begin{abstract}
ժҪA fifth-order nonlinear Schr\"{o}dinger equation which describes one-dimensional anisotropic Heisenberg ferromagnetic spin chain is under exploration in this paper. Starting from the spectral analysis of the Lax pair, a Riemann-Hilbert problem is set up. After solving the obtained Riemann-Hilbert problem with reflectionless case, we systematically derive multi-soliton solutions for the fifth-order nonlinear Schr\"{o}dinger equation. In addition, the localized structures and dynamic behaviors of one- and two-soliton solutions are shown vividly via a few plots. \vskip
2mm\noindent\textbf{AMS Subject classification}: 35C08\vskip
2mm\noindent\textbf{Keywords}: fifth-order nonlinear Schr\"{o}dinger equation, Riemann-Hilbert approach, soliton solutions, localized structures, dynamic behaviors
\end{abstract}

\section{Introduction}
One of the three branches of nonlinear science is the theory of solitons.
Due to the fact that investigations on soliton solutions to nonlinear evolution equations (NLEEs) can provide more insight into interpreting nonlinear phenomena in the fields of hydrodynamics, plasma dynamics, optical communication, solid state physics and so forth, it is a significant work to look for more novel soliton solutions of NLEEs. For this reason, a number of effective approaches have been available thus far for deriving soliton solutions, some of which include the Hirota's bilinear method [1--3], the Darboux transformation method [4--6], the Riemann-Hilbert approach [7--16], the KP hierarchy reduction method [17], the generalized unified method [18]. In recent years, there has been an increasing interest in treating NLEEs with initial-boundary value (IBV) conditions via the Riemann-Hilbert approach. For example, Wang $et\ al.$ [14] studied the Kundu-Eckhaus equation via the Riemann-Hilbert approach from two aspects: first of all, the initial value problem of the defocusing Kundu-Eckhaus equation was considered and its long-time asymptotics was gained. Then the linear spectral problem of the focusing Kundu-Eckhaus equation was investigated through the Riemann-Hilbert formulation and $N$-bright-soliton solution to this equation were found explicitly. More recently, the coupled modified nonlinear Schr\"{o}dinger equations [15] were discussed in the framework of a Riemann-Hilbert problem, based on which $N$-soliton solution for the equations under consideration were generated.

In this paper, we consider a fifth-order nonlinear Schr\"{o}dinger (NLS) equation [19]
\begin{equation}
\begin{aligned}
& i{{q}_{t}}+\frac{1}{2}{{q}_{xx}}+{{\left| q \right|}^{2}}q-i\alpha \big({{q}_{xxx}}+6{{\left| q \right|}^{2}}{{q}_{x}}\big)+\gamma \big({{q}_{xxxx}}+6q_{x}^{2}{{q}^{*}}+4q{{\left| {{q}_{x}} \right|}^{2}}+8{{\left| q \right|}^{2}}{{q}_{xx}}+2{{q}^{2}}q_{xx}^{*}\\
& +6q{{\left| q \right|}^{4}}\big)-i\delta \big({{q}_{xxxxx}}+10{{\left| q \right|}^{2}}{{q}_{xxx}}+30{{\left| q \right|}^{4}}{{q}_{x}}+10q{{q}_{x}}q_{xx}^{*}+10qq_{x}^{*}{{q}_{xx}}+20{{q}^{*}}{{q}_{x}}{{q}_{xx}}\\
& +10q_{x}^{2}q_{x}^{*}\big)=0,
\end{aligned}
\end{equation}
which is used to describe one-dimensional anisotropic Heisenberg ferromagnetic spin chain. Here $q$ represents a normalized complex amplitude of the optical pulse envelope, the subscripts denote the partial derivatives with respect to the scaled spatial coordinate $x$ and time coordinate $t$ correspondingly, whereas $\alpha$, $\gamma$ and $\delta$ are respectively the real coefficients of the third-, fourth- and fifth-order terms, and the asterisk signifies complex conjugation. Actually, eq. (1) covers many significant nonlinear differential equations, which are given below:

(i) When $\alpha=\gamma=\delta=0$, eq. (1) is reduced to the focusing NLS equation [20] describing the wave evolution in different physical systems.

(ii) When $\gamma=\delta=0$ and $\alpha\neq0$, eq. (1) becomes the Hirota equation [21] describing the propagation of a subpicosecond or femtosecond pulse.

(iii) When $\alpha=\delta=0$ and $\gamma\neq0$, eq. (1) is turned into the fourth-order dispersive NLS equation [22] describing the one-dimensional anisotropic Heisenberg ferromagnetic spin chain with the octuple-dipole interaction.

(iv) When $\alpha=\gamma=0$ and $\delta\neq0$, eq. (1) is converted into the fifth-order NLS equation [23] describing the Heisenberg ferromagnetic spin system.

There have been several studies on the fifth-order NLS equation (1) up to now. For instance, the study in Ref. [19] offered the Lax pair, and exact expressions for the most representative soliton solutions, which involves two-soliton collisions and the degenerate case of the two-soliton solution, as well as beating structures composed of two or three solitons, were attained by applying the Darboux transformations.
In another study [24], infinitely-many conservation laws for eq. (1) were found on basis of the Lax pair. Making use of the Hirota's bilinear method, the one-, two- and three-soliton solutions in analytic forms were generated as well. In addition,
the Akhmediev breathers, Kuznetsov-Ma solitons and rogue wave solutions were explored by employing the Darboux transformation method [25].

The current study seeks to find multi-soliton solutions for the fifth-order NLS equation (1) via the Riemann-Hilbert approach.
The outline of the paper is as follows:
in Section 2, we establish a Riemann-Hilbert problem for eq. (1) through performing the spectral analysis of the Lax pair. In Section 3, upon the obtained Riemann-Hilbert problem, we construct multi-soliton solutions for eq. (1). A brief conclusion is given in the last section.

\section{The Riemann-Hilbert problem}
The aim of this section is to formulate a Riemann-Hilbert problem for the fifth-order NLS equation (1), whose Lax pair [19] reads
\begin{subequations}
\begin{align}
 & {{\Phi }_{x}}=U\Phi ,\quad U=i\left( \begin{matrix}
   \zeta  & {{q}^{*}}  \\
   q & -\zeta   \\
\end{matrix} \right),\quad  \\
 & {{\Phi }_{t}}=V\Phi ,\quad V=\sum\limits_{c=0}^{5}{i{{\zeta}^{c}}\left( \begin{matrix}
   {{A}_{c}} & B_{c}^{*}  \\
   {{B}_{c}} & -{{A}_{c}}  \\
\end{matrix} \right)},
\end{align}
\end{subequations}
where $\Phi ={(\varphi(x,t),\psi(x,t))^\textrm{T}}$ is the vector eigenfunction, the symbol $\textrm{T}$ means transpose of the vector, and $\zeta$ is a complex eigenvalue parameter. Besides,
\[\begin{aligned}
 & {{A}_{5}}=16\delta ,\quad {{A}_{4}}=-8\gamma ,\quad {{A}_{3}}=-4\alpha -8\delta {{\left| q  \right|}^{2}},\quad {{A}_{2}}=1+4\gamma {{\left| q  \right|}^{2}}+4i\gamma (q _{x}^{*}q -{q_{x}}{{q}^{*}}), \\
 & {{A}_{1}}=2\alpha {{\left| q \right|}^{2}}+6\delta {{\left| q \right|}^{4}}-2i\gamma (q_{x}^{*}q-{q_{x}}{{q}^{*}})+2\delta (q_{xx}^{*}q-{{\left| {{q}_{x}} \right|}^{2}}+{{q}_{xx}}{{q}^{*}}), \\
 & {{A}_{0}}=-\frac{1}{2}{{\left| q  \right|}^{2}}-3\gamma {{\left| q \right|}^{4}}-i\alpha (q_{x}^{*}q-{{q}_{x}}{{q}^{*}})-\gamma (q_{xx}^{*}q-{{\left| {{q }_{x}} \right|}^{2}}+{{q}_{xx}}{{q}^{*}})-i\delta (q_{xxx}^{*}q \\
 & \quad\quad\ -q_{xx}^{*}{{q}_{x}}+{{q}_{xx}}q_{x}^{*}-{{q}_{xxx}}{{q}^{*}})-6i\delta (q_{x}^{*}q-{{q}_{x}}{{q}^{*}}){{\left| q  \right|}^{2}}, \\
 & {{B}_{5}}=0,\quad {{B}_{4}}=16\delta q,\quad {{B}_{3}}=-8\gamma q +8i\delta {{q}_{x}},\quad {{B}_{2}}=-4\alpha q-8\delta {{\left| q  \right|}^{2}}q-4i\gamma {{q}_{x}}-4\delta {{q}_{xx}},\\
 & {{B}_{1}}=q+4\gamma {{\left| q \right|}^{2}}q-2i\alpha {{q}_{x}}-12i\delta {{\left| q \right|}^{2}}{{q}_{x}}+2\gamma {{q}_{xx}}-2i\delta {{q}_{xxx}}, \\
 & {{B}_{0}}=2\alpha {{\left| q \right|}^{2}}q+6\delta {{\left| q \right|}^{4}}q+\frac{1}{2}i{{q}_{x}}+6i\gamma {{\left| q  \right|}^{2}}{{q}_{x}}+\alpha {{q}_{xx}}+2\delta q_{xx}^{*}{{q}^{2}}+4\delta {{\left| {{q}_{x}} \right|}^{2}}q+6\delta q_{x}^{2}{{q}^{*}} \\
 & \quad\quad\ +8\delta {{q}_{xx}}{{\left| q  \right|}^{2}}+i\gamma {{q}_{xxx}}+\delta {{q}_{xxxx}}.
\end{aligned}\]

For the convenience of analysis,
we write the Lax pair (2) as the equivalent form
\begin{subequations}
\begin{align}
& {{\Phi }_{x}}=i(\zeta \sigma +Q)\Phi ,\\
& {{\Phi }_{t}}=\left( \big(16i\delta {{\zeta}^{5}}-8i\gamma {{\zeta}^{4}}-4i\alpha {{\zeta}^{3}}+i{{\zeta}^{2}}\big)\sigma +\tilde{Q} \right)\Phi ,
\end{align}
\end{subequations}
where
\[\begin{aligned}
  & \sigma =\left( \begin{matrix}
   1 & 0  \\
   0 & -1  \\
\end{matrix} \right),\quad Q=\left( \begin{matrix}
   0 & {{q}^{*}}  \\
   q  & 0  \\
\end{matrix} \right),\end{aligned}\]\[\begin{aligned}
  & \tilde{Q}=i\left( \begin{matrix}
   {{A}_{0}} & B_{0}^{*}  \\
   {{B}_{0}} & -{{A}_{0}}  \\
\end{matrix} \right)+i\zeta \left( \begin{matrix}
   {{A}_{1}} & B_{1}^{*}  \\
   {{B}_{1}} & -{{A}_{1}}  \\
\end{matrix} \right)+i{{\zeta}^{2}}\left( \begin{matrix}
   0 & B_{2}^{*}  \\
   {{B}_{2}} & 0  \\
\end{matrix} \right)+i{{\zeta}^{3}}\left( \begin{matrix}
   0 & B_{3}^{*}  \\
   {{B}_{3}} & 0  \\
\end{matrix} \right)
+i{{\zeta}^{4}}\left( \begin{matrix}
   0 & B_{4}^{*}  \\
   {{B}_{4}} & 0  \\
\end{matrix} \right)\\&\quad\quad+i{{\zeta}^{5}}\left( \begin{matrix}
   0 & B_{5}^{*}  \\
   {{B}_{5}} & 0  \\
\end{matrix} \right)
+i{{\zeta}^{2}}\big[4\gamma {{\left| q \right|}^{2}}+4i\gamma (q_{x}^{*}q-{{q}_{x}}{{q}^{*}})\big]\sigma
-8i{{\zeta}^{3}}\delta {{\left| q \right|}^{2}}\sigma.
\end{aligned}\]

Here we posit that the potential function $q$ in the Lax pair (3) decays to zero sufficiently fast as $x\rightarrow\pm\infty$. It can be seen from (3) that when $x\rightarrow\pm\infty$,
\[\Phi \propto{{\text{e}}^{i\zeta \sigma x+(16i\delta {{\zeta}^{5}}-8i\gamma {{\zeta}^{4}}-4i\alpha {{\zeta}^{3}}+i{{\zeta}^{2}})\sigma t}}.\]
This leads us to introduce the variable transformation
\begin{equation*}
\Phi =\mu{{\text{e}}^{i\zeta \sigma x+(16i\delta {{\zeta}^{5}}-8i\gamma {{\zeta}^{4}}-4i\alpha {{\zeta}^{3}}+i{{\zeta}^{2}})\sigma t}},
\end{equation*}
based on which the Lax pair (3) is then converted into
\begin{subequations}
\begin{align}
& {{\mu}_{x}}=i\zeta[\sigma ,\mu]+{{U}_{1}}\mu,\\
& {{\mu}_{t}}=\big(16i\delta {{\zeta}^{5}}-8i\gamma {{\zeta}^{4}}-4i\alpha {{\zeta}^{3}}+i{{\zeta}^{2}}\big)[\sigma ,\mu]+\tilde{Q}\mu,
\end{align}
\end{subequations}
where $[\sigma,\mu]\equiv\sigma \mu-\mu\sigma $ is the commutator and ${{U}_{1}}=iQ$.

Now we begin to consider the spectral analysis of the Lax pair (4), for which we merely concentrate on the spectral problem (4a).
Because the analysis will take place at a fixed time, the $t$-dependence will be suppressed.
As for (4a), we can write its two matrix Jost solutions as a collection of columns, that is
\begin{equation}
{{\mu}_{\pm}}=({{[{{\mu}_{\pm}}]}_{1}},{{[{{\mu}_{\pm}}]}_{2}})
\end{equation}
obeying the asymptotic conditions
\begin{subequations}
\begin{align}
&{{\mu}_{-}}\to \mathbb{I},\quad x\to -\infty ,\\
&{{\mu}_{+}}\to \mathbb{I},\quad x\to +\infty .
\end{align}
\end{subequations}
Here the subscripts of $\mu$ signify which end of the $x$-axis the boundary conditions are set, and $\mathbb{I}$ is a $2\times2$ unit matrix. The ${{\mu}_{\pm }}$ are uniquely determined by the integral equations of Volterra type
\begin{subequations}
\begin{align}
&{{\mu}_{-}}=\mathbb{I}+\int_{-\infty }^{x}{{{\text{e}}^{i\zeta \sigma (x-y)}}{{U}_{1}}(y){{\mu}_{-}}(y,\zeta){{\text{e}}^{-i\zeta \sigma (x-y )}}\text{d}y},\\
&{{\mu}_{+}}=\mathbb{I}-\int_{x}^{+\infty }{{{\text{e}}^{i\zeta \sigma (x-y)}}{{U}_{1}}(y){{\mu}_{+}}(y,\zeta){{\text{e}}^{-i\zeta \sigma (x-y )}}\text{d}y}.
\end{align}
\end{subequations}
By the direct analysis of (7), we can see that ${{[{{\mu}_{-}}]}_{1}},{{[{{\mu}_{+}}]}_{2}}$ are analytic for $\zeta \in {\mathbb{D}^{-}}$ and continuous for $\zeta \in {\mathbb{D}^{-}}\cup \mathbb{R}$, while ${{[{{\mu}_{+}}]}_{1}},{{[{{\mu}_{-}}]}_{2}}$ are analytic for $\zeta \in {\mathbb{D}^{+}}$ and continuous for $\zeta \in {\mathbb{D}^{+}}\cup \mathbb{R}$, where ${\mathbb{D}^{-}}$ and ${\mathbb{D}^{+}}$ are respectively the lower and upper half $\zeta$-plane.

It is indicated owing to the Abel's identity and $\text{tr}Q=0$ that the determinants of ${{\mu}_{\pm }}$ are independent of the variable $x$. Through evaluating $\det {{\mu}_{-}}$ at $x=-\infty$ and $\det {{\mu}_{+}}$ at $x=+\infty$, we know $\det {{\mu}_{\pm }}=1$ for $ \zeta \in \mathbb{R}.$
Since both ${{\mu}_{-}}E$ and ${{\mu}_{+}}E$ are matrix solutions of the spectral problem (4a),
where $E={{\text{e}}^{i\zeta \sigma x}}$,
they must be linearly associated, namely
\begin{equation}
{{\mu}_{-}}E={{\mu}_{+}}ES(\zeta),\quad \zeta \in \mathbb{R},
\end{equation}
with $S(\zeta )={{({{s}_{kj}})}_{2\times 2}}$
being a scattering matrix.
It is obvious that
$
\det S(\zeta)=1.
$
Furthermore, we can see from the property of $\mu_{-}$ that ${{s}_{11}}$ allows analytic extension to ${\mathbb{D}^{+}}$ and ${{s}_{kj}}$\ $(k,j=1,2)$ analytically extend to ${\mathbb{D}^{-}}$.

A Riemann-Hilbert problem we are now looking for is related to two matrix functions: one is analytic in ${\mathbb{D}^{+}}$ and the other is analytic in ${\mathbb{D}^{-}}$. Set the first analytic function of $\zeta$ in ${\mathbb{D}^{+}}$ be
\begin{equation}
{{P}_{1}}(x,\zeta)=({{[{{\mu}_{+}}]}_{1}},{{[{{\mu}_{-}}]}_{2}})(x,\zeta).
\end{equation}
And then, ${{P}_{1}}$ can be expanded into the asymptotic series at large-$\zeta$
\begin{equation}
{{P}_{1}}=P_{1}^{(0)}+\frac{P_{1}^{(1)}}{\zeta}+\frac{P_{1}^{(2)}}{{{\zeta}^{2}}}+O\bigg( \frac{1}{{{\zeta}^{3}}} \bigg),\quad \zeta \to \infty .
\end{equation}
Since ${{P}_{1}}$ satisfies the spectral problem (4a), inserting (10) into (4a) and equating terms with like powers of $\zeta$ directly brings about
\begin{align}
& O(1):i\big[\sigma ,P_{1}^{(1)}\big]+{{U}_{1}}P_{1}^{(0)}=P_{1x}^{(0)},\nonumber\\
& O(\zeta):i\big[\sigma ,P_{1}^{(0)}\big]=0, \nonumber
\end{align}
from which we have $P_{1}^{(0)}=\mathbb{I}$, namely
${{P}_{1}}\to \mathbb{I}$ as $ \lambda \in {\mathbb{D}^{+}}\to \infty .$

In order to work out a Riemann-Hilbert problem, there is still the analytic counterpart of $P_{1}$ in ${\mathbb{D}^{-}}$ to be found.
For this purpose, we partition the inverse matrices of ${{\mu}_{\pm }}$ into rows, that is
\begin{equation}
\mu_{\pm}^{-1}=\left( \begin{matrix}
   {[\mu_{\pm}^{-1}]^{1}}  \\
   {[\mu_{\pm}^{-1}]^{2}}  \\
\end{matrix} \right),
\end{equation}
which fulfill the adjoint scattering equation relating to (4a)
\begin{equation}
{{K}_{x}}=i\zeta[\sigma ,K]-K{{U}_{1}},
\end{equation}
and follow the boundary conditions $\mu_{\pm}^{-1}\rightarrow\mathbb{I}$ as $x\rightarrow\pm\infty$.
It is apparent from (8) that
\begin{equation}
{{E}^{-1}}\mu_{-}^{-1}=R(\zeta){{E}^{-1}}\mu_{+}^{-1},
\end{equation}
where $R(\zeta)={{({{r}_{kj}})}_{2\times 2}}$ is the inverse matrix of $S(\zeta)$.
Hence, the matrix function ${{P}_{2}}$ which is analytic in ${\mathbb{D}^{-}}$ is of the form
\begin{equation}
{{P}_{2}}(x,\zeta)=\left( \begin{matrix}
   {[\mu_{+}^{-1}]^{1}}  \\
   {[\mu_{-}^{-1}]^{2}}  \\
\end{matrix} \right)(x,\zeta).
\end{equation}
In the same way as ${{P}_{1}}$, it turns out that the very large-$\zeta$ asymptotic behavior of ${{P}_{2}}$ is
$
{{P}_{2}}\to \mathbb{I}$ as $\zeta \in {\mathbb{D}^{-}}\to \infty .
$

Substituting (5) into (8) gives rise to
\begin{equation*}
({{[{{\mu}_{-}}]}_{1}},{{[{{\mu}_{-}}]}_{2}})=({{[{{\mu}_{+}}]}_{1}},{{[{{\mu}_{+}}]}_{2}})\left( \begin{matrix}
   {{s}_{11}} & {{s}_{12}}{{\text{e}}^{2i\zeta x}}  \\
   {{s}_{21}}{{\text{e}}^{-2i\zeta x}} & {{s}_{22}}  \\
\end{matrix} \right),
\end{equation*}
from which we obtain
\begin{equation*}
{{[{{\mu}_{-}}]}_{2}}={{s}_{12}}{{\text{e}}^{2i\zeta x}}{{[{{\mu}_{+}}]}_{1}}+{{s}_{22}}{{[{{\mu}_{+}}]}_{2}}.
\end{equation*}
Hence, ${{P}_{1}}$ is of the form
\begin{equation*}
{{P}_{1}}=({{[{{\mu}_{+}}]}_{1}},{{[{{\mu}_{-}}]}_{2}})=({{[{{\mu}_{+}}]}_{1}},{{[{{\mu}_{+}}]}_{2}})\left( \begin{matrix}
   1 & {{s}_{12}}{{\text{e}}^{2i\zeta x}}  \\
   0 & {{s}_{22}}  \\
\end{matrix} \right).
\end{equation*}

On the other hand, via carrying (11) into (13), we derive
\begin{equation*}
\left( \begin{matrix}
   {[\mu_{-}^{-1}]^{1}}  \\
   {[\mu_{-}^{-1}]^{2}}  \\
\end{matrix} \right)=\left( \begin{matrix}
   {{r}_{11}} & {{r}_{12}}{{\text{e}}^{2i\zeta x}}  \\
   {{r}_{21}}{{\text{e}}^{-2i\zeta x}} & {{r}_{22}}  \\
\end{matrix} \right)\left( \begin{matrix}
   {[\mu_{+}^{-1}]^{1}}  \\
   {[\mu_{+}^{-1}]^{2}}  \\
\end{matrix} \right),
\end{equation*}
from which we can express ${[\mu_{-}^{-1}]^{2}}$ as
\begin{equation*}
{[\mu_{-}^{-1}]^{2}}={{r}_{21}}{{\text{e}}^{-2i\zeta x}}{[\mu_{+}^{-1}]^{1}}+{{r}_{22}}{[\mu_{+}^{-1}]^{2}}.
\end{equation*}
Subsequently, ${{P}_{2}}$ is represented as
\begin{equation*}
{{P}_{2}}=\left( \begin{matrix}
   {[\mu_{+}^{-1}]^{1}}  \\
   {[\mu_{-}^{-1}]^{2}}  \\
\end{matrix} \right)=\left( \begin{matrix}
   1 & 0  \\
   {{r}_{21}}{{\text{e}}^{-2i\zeta x}} & {{r}_{22}}  \\
\end{matrix} \right)\left( \begin{matrix}
   {[\mu_{+}^{-1}]^{1}}  \\
   {[\mu_{+}^{-1}]^{2}}  \\
\end{matrix} \right).
\end{equation*}

Having found two matrix functions ${{P}_{1}}$ and ${{P}_{2}}$ which are analytic in ${\mathbb{D}^{+}}$ and ${\mathbb{D}^{-}}$ respectively so far, we are in a position to set up a Riemann-Hilbert problem for the fifth-order NLS equation (1).
Here we denote that the limit of ${{P}_{1}}$ is ${{P}^{+}}$ as $\zeta \in {\mathbb{D}^{+}}\rightarrow\mathbb{R}$ and
the limit of ${{P}_{2}}$ is ${{P}^{-}}$ as $\zeta \in {\mathbb{D}^{-}}\rightarrow\mathbb{R}$, based on which a Riemann-Hilbert problem desired can be got below
\begin{equation}
{{P}^{-}}(x,\zeta){{P}^{+}}(x,\zeta)=\left( \begin{matrix}
   1 & {{s}_{12}}{{\text{e}}^{2i\zeta x}}  \\
   {{r}_{21}}{{\text{e}}^{-2i\zeta x}} & 1  \\
\end{matrix} \right),
\end{equation}
whose canonical normalization conditions are
\begin{eqnarray*}
  &{{P}_{1}}(x,\zeta)\to \mathbb{I},\quad \zeta \in {\mathbb{D}^{+}}\to \infty , \\
  &{{P}_{2}}(x,\zeta)\to \mathbb{I},\quad \zeta \in {\mathbb{D}^{-}}\to \infty ,
\end{eqnarray*}
and ${{r}_{21}}{{s}_{12}}+{{r}_{22}}{{s}_{22}}=1$.

In what follows, we plan to retrieve the potential function $q(x,t)$. As a matter of fact, expanding ${{P}_{1}}(\zeta)$ at large-$\zeta$ as
\begin{equation*}
{{P}_{1}}(\zeta)=\mathbb{I}+\frac{P_{1}^{(1)}}{\zeta}+\frac{P_{1}^{(2)}}{{{\zeta}^{2}}}+O\bigg( \frac{1}{{{\zeta}^{3}}} \bigg),\quad \zeta \to \infty .
\end{equation*}
and then inserting this expansion into (4a) yields
\begin{equation*}
Q=-\big[\sigma ,P_{1}^{(1)}\big].
\end{equation*}
Thereupon, the potential function is restructured as
\begin{equation*}
q(x,t)=2{{\big(P_{1}^{(1)}\big)}_{21}},
\end{equation*}
where ${{\big(P_{1}^{(1)}\big)}_{21}}$ is the (2,1)-entry of matrix $P_{1}^{(1)}$.

\section{Multi-soliton solutions}
The previous section has described a Riemann-Hilbert problem for the fifth-order NLS equation (1). Next we intend to solve the Riemann-Hilbert problem in the sense of irregularity.
The irregularity means that both $\det {{P}_{1}}$ and $\det {{P}_{2}}$ possess some zeros in their analytic domains.
By drawing on the definitions of ${{P}_{1}}$ and ${{P}_{2}}$, we have
\begin{align}
& \det {{P}_{1}}(\zeta)={{s}_{22}}(\zeta),\quad \zeta \in {\mathbb{D}^{+}},  \nonumber \\
& \det {{P}_{2}}(\zeta)={{r}_{22}}(\zeta),\quad \zeta \in {\mathbb{D}^{-}},  \nonumber
\end{align}
from which we could find that
$\det {{P}_{1}}$ and $\det {{P}_{2}}$ possess the same zeros as ${s}_{22}$ and ${r}_{22}$ respectively, and ${{r}_{22}}={{({{S}^{-1}})}_{22}}={{s}_{11}}$.

In view of the above, it is time to reveal the characteristic feature of zeros.
Regarding the potential matrix $Q$ having the symmetry relation
$
Q^{\dagger }=Q,
$
where the superscript $\dagger$ signifies the Hermitian of a matrix, we conclude
\begin{equation}
\mu_{\pm }^{\dagger }({{\zeta}^{*}})=\mu_{\pm }^{-1}(\zeta).
\end{equation}
In order to facilitate discussion, we express (9) and (14) in terms of
\begin{subequations}
\begin{align}
&{{P}_{1}}={{\mu}_{+}}{{H}_{1}}+{{\mu}_{-}}{{H}_{2}},\\
&{{P}_{2}}={{H}_{1}}\mu_{+}^{-1}+{{H}_{2}}\mu_{-}^{-1},
\end{align}
\end{subequations}
where ${{H}_{1}}=\text{diag}(1,0)$ and ${{H}_{2}}=\text{diag}(0,1).$
Taking the Hermitian of expression (17a) and making use of the relation (16), we find
\begin{equation}
P_{1}^{\dagger }({{\zeta}^{*}})={{P}_{2}}(\zeta),\quad \zeta \in {\mathbb{D}^{-}},
\end{equation}
and the involution property of the scattering matrix
$
{{S}^{\dagger }}({{\zeta}^{*}})={{S}^{-1}}(\zeta).
$
This evidently follows
\begin{equation}
s_{22}^{*}({{\zeta}^{*}})={{r}_{22}}(\zeta),\quad \zeta \in {\mathbb{D}^{-}},
\end{equation}
which suggests that each zero $\pm {{\zeta}_{k}}$ of ${{s}_{22}}$ leads to each zero $\pm \zeta_{k}^{*}$ of ${{r}_{22}}$ correspondingly.
Therefore, we suppose that $\det {{P}_{1}}$ has $N$ simple zeros $\{{{\zeta }_{j}}\}_{1}^{N}$ in ${\mathbb{D}^{+}}$
and $\det {{P}_{2}}$ has $N$ simple zeros $\{{{\hat{\zeta }}_{j}}\}_{1}^{N}$ in ${\mathbb{D}^{-}}$, where
${{\hat{\zeta}}_{l}}=\zeta_{l}^{*},$ $ 1\le l\le N.$
These zeros together with the nonzero vectors ${{\nu}_{j}}$ and ${{\hat{\nu}}_{j}}$ constitute the full set of the generic discrete data, which meet the equations
\begin{subequations}
\begin{align}
&{{P}_{1}}({{\zeta}_{j}}){{\nu}_{j}}=0,\\
&{{\hat{\nu}}_{j}}{{P}_{2}}({{\hat{\zeta}}_{j}})=0,
\end{align}
\end{subequations}
where ${{\nu}_{j}}$ and ${{\hat{\nu}}_{j}}$ denote column vectors and row vectors respectively.
Taking the Hermitian of eq. (20a) and using (18) as well as comparing with eq. (20b), we know that the eigenvectors fulfill the relation
\begin{equation}
{{\hat{\nu}}_{j}}=\nu_{j}^{\dagger },\quad 1\le j\le N.
\end{equation}
By differentiations of eq. (20a) with respect to $x$ and $t$ respectively and use of the Lax pair (4), we get
\begin{equation*}
\begin{aligned}
 & {{P}_{1}}({{\zeta }_{j}})\left( \frac{\partial {{\nu }_{j}}}{\partial x}-i{{\zeta }_{j}}\sigma {{\nu }_{j}} \right)=0, \\
 & {{P}_{1}}({{\zeta }_{j}})\left( \frac{\partial {{\nu }_{j}}}{\partial t}-\left( 16i\delta \zeta _{j}^{5}-8i\gamma \zeta _{j}^{4}-4i\alpha \zeta _{j}^{3}+i\zeta _{j}^{2}\right)\sigma {{\nu }_{j}} \right)=0,
\end{aligned}
\end{equation*}
which generates
\begin{equation*}
{{\nu}_{j}}\text{=}{{\text{e}}^{(i{{\zeta}_{j}}x+(16i\delta \zeta _{j}^{5}-8i\gamma \zeta_{j}^{4}-4i\alpha \zeta_{j}^{3}+i\zeta_{j}^{2})t)\sigma }}{{\nu}_{j,0}},\quad 1\le j\le N.
\end{equation*}
Here ${{\nu}_{j,0}}$ are complex constant vectors. Also from the relation (21) we have
\[{{\hat{\nu}}_{j}}=\nu_{j,0}^{\dagger }{{\text{e}}^{(-i\zeta _{j}^{*}x+(-16i\delta \zeta {{_{j}^{*}}^{5}}+8i\gamma \zeta {{_{j}^{*}}^{4}}+4i\alpha \zeta {{_{j}^{*}}^{3}}-i\zeta {{_{j}^{*}}^{2}})t)\sigma}}.\]

Worthy to note that, the Riemann-Hilbert problem (15) we treat corresponds to reflectionless case, i.e. ${{s}_{12}}=0$.
The solution for the Riemann-Hilbert problem (15) can be given as
\begin{subequations}
\begin{align}
& {{P}_{1}}(\zeta)=\mathbb{I}-\sum\limits_{k=1}^{N}{\sum\limits_{j=1}^{N}{\frac{{{\nu}_{k}}{{{\hat{\nu}}}_{j}}{{\big({{M}^{-1}}\big)}_{kj}}}{\zeta -{{{\hat{\zeta}}}_{j}}}}}, \\
& {{P}_{2}}(\zeta)=\mathbb{I}+\sum\limits_{k=1}^{N}{\sum\limits_{j=1}^{N}{\frac{{{\nu}_{k}}{{{\hat{\nu}}}_{j}}{{\big({{M}^{-1}}\big)}_{kj}}}{\zeta-{{\zeta }_{k}}}}},
\end{align}
\end{subequations}
where $M$ is a $N\times N$ matrix with entries
\begin{equation*}
{{m}_{kj}}=\frac{{\hat{\nu}_{k}}{{{{\nu}}}_{j}}}{{{\zeta}_{j}}-{{{\hat{\zeta}}}_{k}}},\quad 1\le k,j\le N,
\end{equation*}
and ${{\big({{M}^{-1}}\big)}_{kj}}$ means the $(k,j)$-entry of the inverse matrix of $M$. From expression (22a), we can immediately obtain
\begin{equation*}
P_{1}^{(1)}=-\sum\limits_{k=1}^{N}{\sum\limits_{j=1}^{N}{{{\nu}_{k}}{{{\hat{\nu}}}_{j}}{{\big({{M}^{-1}}\big)}_{kj}}}}.
\end{equation*}

As a consequence, the expression of general $N$-soliton solution for the fifth-order NLS equation (1) can be derived as follows
\begin{equation}
q=-2\sum\limits_{k=1}^{N}{\sum\limits_{j=1}^{N}{\alpha _{j}^{*}{{\beta }_{k}}{{\text{e}}^{-{{\theta }_{k}}+\theta _{j}^{*}}}{{\big({{M}^{-1}}\big)}_{kj}}}},
\end{equation}
where
\begin{equation*}
{{m}_{kj}}=\frac{\alpha _{k}^{*}{{\alpha }_{j}}{{\text{e}}^{\theta _{k}^{*}+{{\theta }_{j}}}}+\beta _{k}^{*}{{\beta }_{j}}{{\text{e}}^{-\theta _{k}^{*}-{{\theta }_{j}}}}}{{{\zeta}_{j}}-\zeta _{k}^{*}},\quad 1\le k,j\le N.
\end{equation*}
Here we have set nonzero vectors ${{\nu}_{k,0}}={({{\alpha }_{k}},{{\beta }_{k}})^\textrm{T}}$ and ${{\theta }_{k}}=i{{\zeta}_{k}}x+(16i\delta \zeta _{k}^{5}-8i\gamma \zeta_{k}^{4}-4i\alpha \zeta_{k}^{3}+i\zeta_{k}^{2})t$, $(\operatorname{Im}{{\zeta }_{k}}>0,1\le k\le N).$

In the rest of this section, we write out one- and two-bright-soliton solutions explicitly and plot them. In the case of $N=1$, the one-bright-soliton solution can be readily obtained as
\begin{equation}
q=-2\alpha _{1}^{*}{{\beta }_{1}}\frac{({{\zeta}_{1}}-\zeta_{1}^{*}){{\text{e}}^{-{{\theta }_{1}}+\theta _{1}^{*}}}}{{{\left| {{\alpha }_{1}} \right|}^{2}}{{\text{e}}^{\theta _{1}^{*}+{{\theta }_{1}}}}+{{\left| {{\beta }_{1}} \right|}^{2}}{{\text{e}}^{-\theta _{1}^{*}-{{\theta }_{1}}}}},
\end{equation}
in which
${{\theta }_{1}}=i{{\zeta}_{1}}x+(16i\delta \zeta_{1}^{5}-8i\gamma \zeta_{1}^{4}-4i\alpha \zeta_{1}^{3}+i\zeta _{1}^{2})t$.
Furthermore, fixing ${{\beta }_{1}}=1$ and setting ${{\zeta}_{1}}={{a}_{1}}+i{{b}_{1}}$ as well as ${{\left| {{\alpha }_{1}} \right|}^{2}}={{\text{e}}^{2{{\xi }_{1}}}}$, the expression (24) is then turned into
\begin{equation}
q=-2i\alpha _{1}^{*}{{b}_{1}}{{\text{e}}^{-{{\xi }_{1}}}}{{\text{e}}^{\theta _{1}^{*}-{{\theta }_{1}}}}\text{sech}({{\theta }_{1}}+\theta _{1}^{*}+{{\xi }_{1}}).
\end{equation}
According to the notation above, we arrive at
\begin{align*}
& \theta _{1}^{*}+{\theta }_{1}=-2{b}_{1}\big[ x+(80\delta a_{1}^{4}-160\delta a_{1}^{2}b_{1}^{2}+16\delta b_{1}^{4}-32\gamma a_{1}^{3}+32\gamma {{a}_{1}}b_{1}^{2}-12\alpha a_{1}^{2}+4\alpha b_{1}^{2}+2{{a}_{1}})t\big],\\
& \theta _{1}^{*}-{\theta }_{1}=-2ia_{{1}}x+(-160i\delta a_{{1}}b_{1}^{4}+320i\delta
a_{1}^{3}b_{1}^{2}-96i\gamma a_{1}^{2}b_{1}^{2}-24i\alpha a_{{1}}b_{1}^{2}+16i\gamma a_{1}^{4}-32i\delta a_{1}^{5}\\
&\quad\quad\quad\quad\ +8i\alpha a_{1}^{3}+16i\gamma b_{1}^{4}+2ib_{1}^{2}-2ia_{1}^{2})t.
\end{align*}
Thus the one-bright-soliton solution (25) can be further written as
\begin{equation}
\begin{aligned}
q=&-2i\alpha _{1}^{*}{{b}_{1}}{{\text{e}}^{-{{\xi }_{1}}}}{{\text{e}}^{\theta _{1}^{*}-{{\theta }_{1}}}}\text{sech}\big\{-2{b}_{1}\big[ x+(80\delta a_{1}^{4}
-160\delta a_{1}^{2}b_{1}^{2}+16\delta b_{1}^{4}-32\gamma a_{1}^{3}+32\gamma {{a}_{1}}b_{1}^{2}\\
&-12\alpha a_{1}^{2}+4\alpha b_{1}^{2}+2{{a}_{1}})t \big]+{{\xi }_{1}}\big\}.
\end{aligned}
\end{equation}

From expression (26) it can be seen that the one-bright-soliton solution is of the shape of hyperbolic secant function with peak amplitude
\[
\mathcal{A}=2\left| \alpha _{1}^{*} \right|{{b}_{1}}{{\text{e}}^{-{{\xi }_{1}}}},\]
and its velocity
\begin{equation*}
\mathcal{V}=-80\delta a_{1}^{4}+160\delta a_{1}^{2}b_{1}^{2}-16\delta b_{1}^{4}
+32\gamma a_{1}^{3}-32\gamma {{a}_{1}}b_{1}^{2}+12\alpha a_{1}^{2}-4\alpha b_{1}^{2}-2{{a}_{1}}
\end{equation*}
relying on both the real part ${a}_{1}$ and the imaginary part ${b}_{1}$ of the eigenvalue ${\zeta}_{1}$.
The localized structures and dynamic behaviors of the one-bright-soliton solution (26) are depicted in Figures 1, 2 and 3 with parameters chosen as ${{a}_{1}}=0.2,{{b}_{1}}=0.3,{{\xi }_{1}}=0,\alpha =\gamma =\delta ={{\alpha }_{1}}=1$.

In addition,
the two-bright-soliton solution to eq. (1) can be generated by taking $N=2$ in the formula (23)
\begin{equation}
\begin{aligned}
q=&-\frac{2}{{{m}_{11}}{{m}_{22}}-{{m}_{12}}{{m}_{21}}}\big( \alpha _{1}^{*}{{\beta }_{1}}{{m}_{22}}{{\text{e}}^{-{{\theta }_{1}}+\theta _{1}^{*}}}-\alpha _{2}^{*}{{\beta }_{1}}{{m}_{12}}{{\text{e}}^{-{{\theta }_{1}}+\theta _{2}^{*}}}\\
&-\alpha _{1}^{*}{{\beta }_{2}}{{m}_{21}}{{\text{e}}^{-{{\theta }_{2}}+\theta _{1}^{*}}}+\alpha _{2}^{*}{{\beta }_{2}}{{m}_{11}}{{\text{e}}^{-{{\theta }_{2}}+\theta _{2}^{*}}}\big),
\end{aligned}
\end{equation}
where
\[\begin{aligned}
 & {{m}_{11}}=\frac{{{\left| {{\alpha }_{1}} \right|}^{2}}{{\text{e}}^{\theta _{1}^{*}+{{\theta }_{1}}}}+{{\left| {{\beta }_{1}} \right|}^{2}}{{\text{e}}^{-\theta _{1}^{*}-{{\theta }_{1}}}}}{{{\zeta}_{1}}-\zeta_{1}^{*}},\quad {{m}_{12}}=\frac{\alpha _{1}^{*}{{\alpha }_{2}}{{\text{e}}^{\theta _{1}^{*}+{{\theta }_{2}}}}+\beta _{1}^{*}{{\beta }_{2}}{{\text{e}}^{-\theta _{1}^{*}-{{\theta }_{2}}}}}{{{\zeta}_{2}}-\zeta _{1}^{*}}, \\
 & {{m}_{21}}=\frac{\alpha _{2}^{*}{{\alpha }_{1}}{{\text{e}}^{\theta _{2}^{*}+{{\theta }_{1}}}}+\beta _{2}^{*}{{\beta }_{1}}{{\text{e}}^{-\theta _{2}^{*}-{{\theta }_{1}}}}}{{{\zeta}_{1}}-\zeta_{2}^{*}},\quad
{{m}_{22}}=\frac{{{\left| {{\alpha }_{2}} \right|}^{2}}{{\text{e}}^{\theta _{2}^{*}+{{\theta }_{2}}}}+{{\left| {{\beta }_{2}} \right|}^{2}}{{\text{e}}^{-\theta _{2}^{*}-{{\theta }_{2}}}}}{{{\zeta}_{2}}-\zeta_{2}^{*}},\end{aligned}\]\[
 {{\theta }_{1}}=i{{\zeta}_{1}}x+(16i\delta \zeta_{1}^{5}-8i\gamma \zeta _{1}^{4}-4i\alpha \zeta _{1}^{3}+i\zeta _{1}^{2})t, \quad
 {{\theta }_{2}}=i{{\zeta}_{2}}x+(16i\delta \zeta_{2}^{5}-8i\gamma \zeta _{2}^{4}-4i\alpha \zeta _{2}^{3}+i\zeta _{2}^{2})t,
\]
and ${{\zeta}_{1}}={{a}_{1}}+i{{b}_{1}},{{\zeta}_{2}}={{a}_{2}}+i{{b}_{2}}.$

If we let ${{\beta }_{1}}={{\beta }_{2}}=1,{{\alpha }_{1}}={{\alpha }_{2}}$ and ${{\left| {{\alpha }_{1}} \right|}^{2}}={{\text{e}}^{2{{\xi }_{1}}}}$, then the two-bright-soliton solution (27) has the form
\begin{equation}
q=-\frac{2}{{{m}_{11}}{{m}_{22}}-{{m}_{12}}{{m}_{21}}}\big( \alpha _{1}^{*}{{m}_{22}}{{\text{e}}^{-{{\theta }_{1}}+\theta _{1}^{*}}}
-\alpha _{2}^{*}{{m}_{12}}{{\text{e}}^{-{{\theta }_{1}}+\theta _{2}^{*}}}-\alpha _{1}^{*}{{m}_{21}}{{\text{e}}^{-{{\theta }_{2}}+\theta _{1}^{*}}}
+\alpha _{2}^{*}{{m}_{11}}{{\text{e}}^{-{{\theta }_{2}}+\theta _{2}^{*}}}\big),
\end{equation}
where
\begin{equation*}
\begin{aligned}
 &{{m}_{11}}=-\frac{i}{{{b}_{1}}}{{\text{e}}^{{{\xi }_{1}}}}\cosh (\theta _{1}^{*}+{{\theta }_{1}}+{{\xi }_{1}}),\quad {{m}_{12}}=\frac{2{{\text{e}}^{{{\xi }_{1}}}}}{({{a}_{2}}-{{a}_{1}})+i({{b}_{1}}+{{b}_{2}})}\cosh (\theta _{1}^{*}+{{\theta }_{2}}+{{\xi }_{1}}), \\
 &{{m}_{21}}=\frac{2{{\text{e}}^{{{\xi }_{2}}}}}{({{a}_{1}}-{{a}_{2}})+i({{b}_{1}}+{{b}_{2}})}\cosh (\theta _{2}^{*}+{{\theta }_{1}}+{{\xi }_{2}}),\quad {{m}_{22}}=-\frac{i}{{{b}_{2}}}{{\text{e}}^{{{\xi }_{2}}}}\cosh (\theta _{2}^{*}+{{\theta }_{2}}+{{\xi }_{2}}).
\end{aligned}
\end{equation*}

\begin{figure}
\begin{center}
\subfigure[]{\resizebox{0.3\hsize}{!}{\includegraphics*{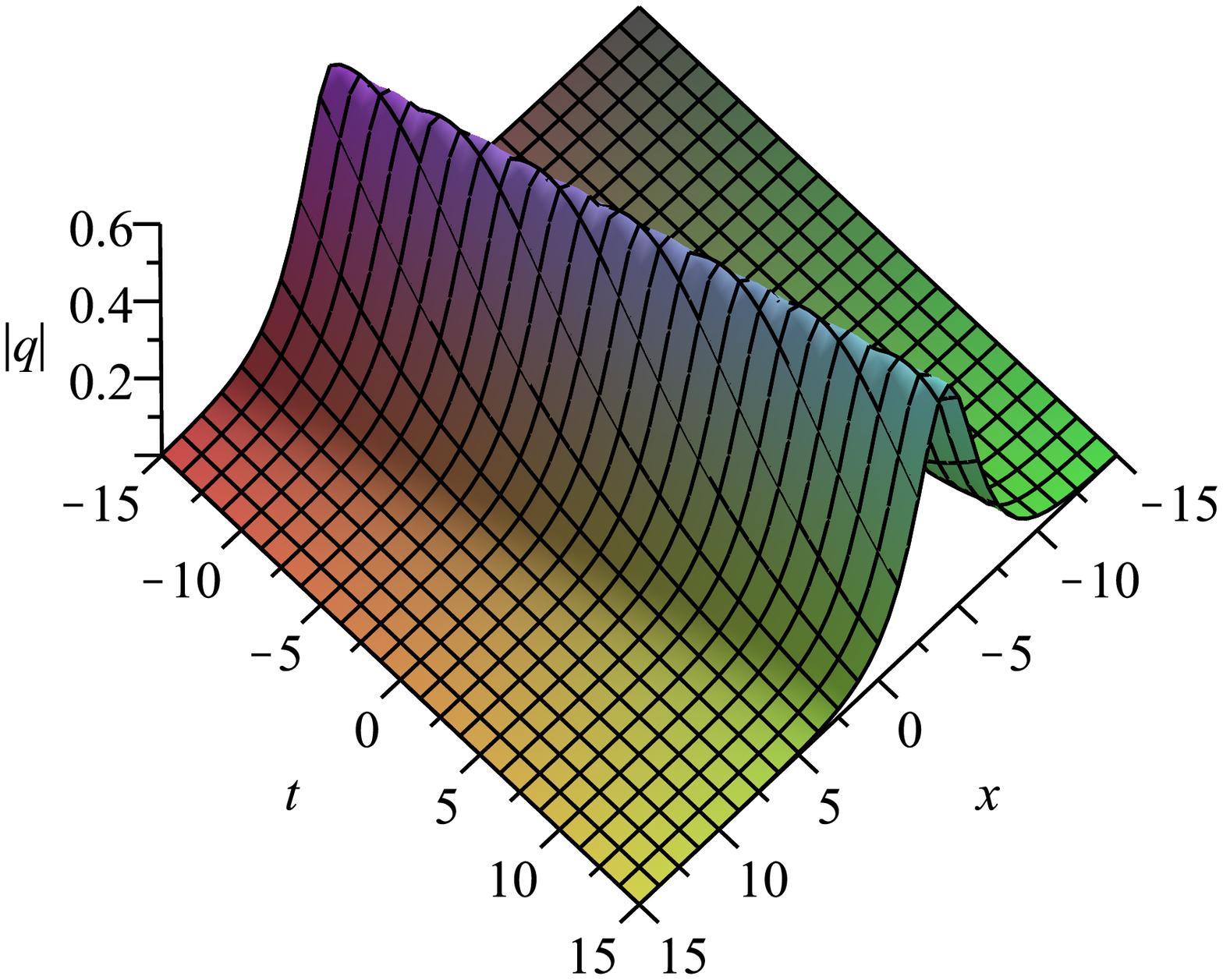}}}
\subfigure[]{\resizebox{0.28\hsize}{!}{\includegraphics*{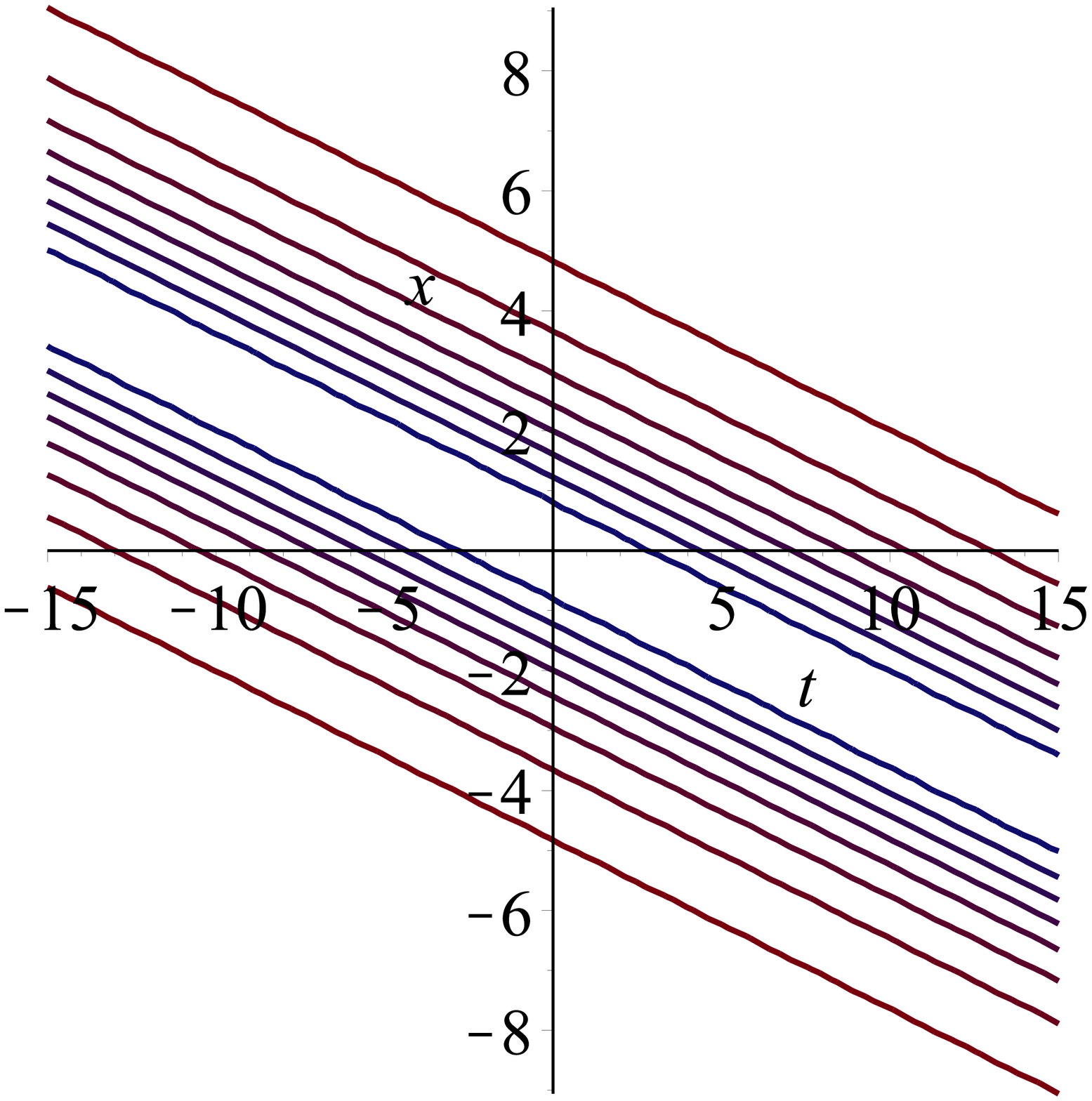}}}
\subfigure[]{\resizebox{0.28\hsize}{!}{\includegraphics*{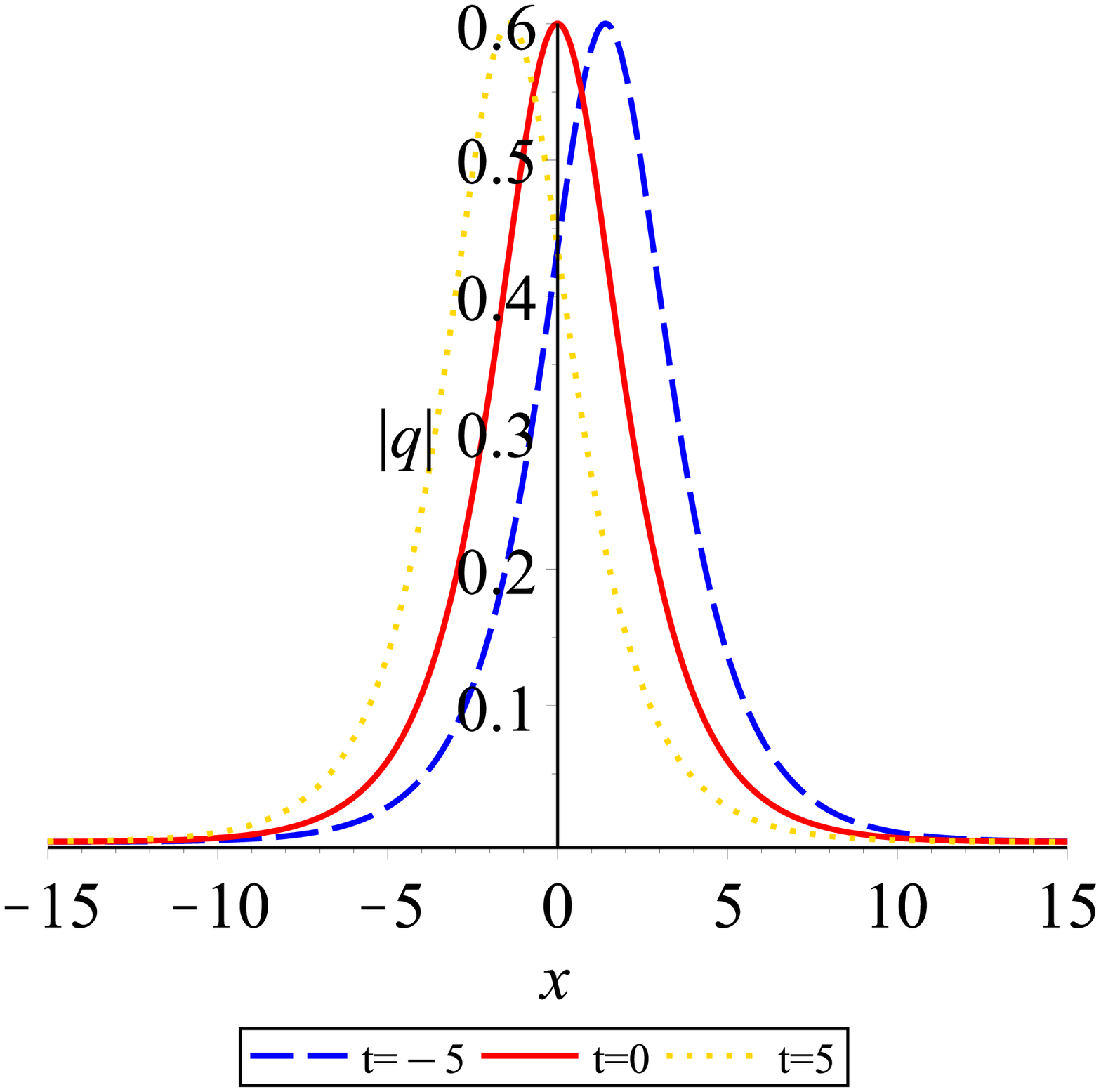}}}
\parbox[c]{13.0cm}{\footnotesize
{\bf Figure 1.}~Plots of one-soliton solution (26). (a) Perspective view of modulus of $q$; (b) Contour plot of modulus of $q$ in Figure 1(a); (c) The soliton along the $x$-axis with different time in Figure 1(a).}
\end{center}
\addtocounter{subfigure}{-3}
\end{figure}
\begin{figure}
\begin{center}
\subfigure[]{\resizebox{0.3\hsize}{!}{\includegraphics*{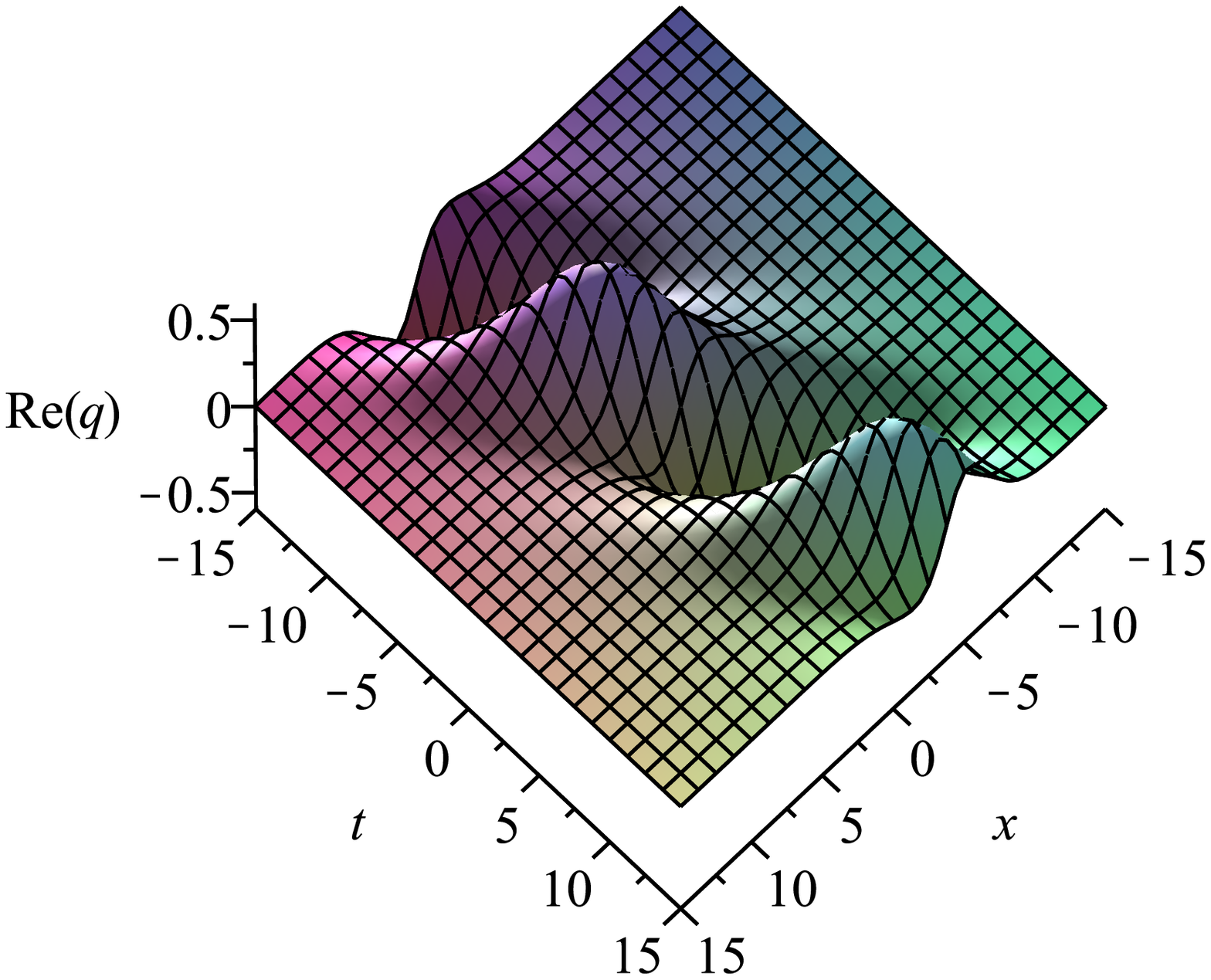}}}
\subfigure[]{\resizebox{0.28\hsize}{!}{\includegraphics*{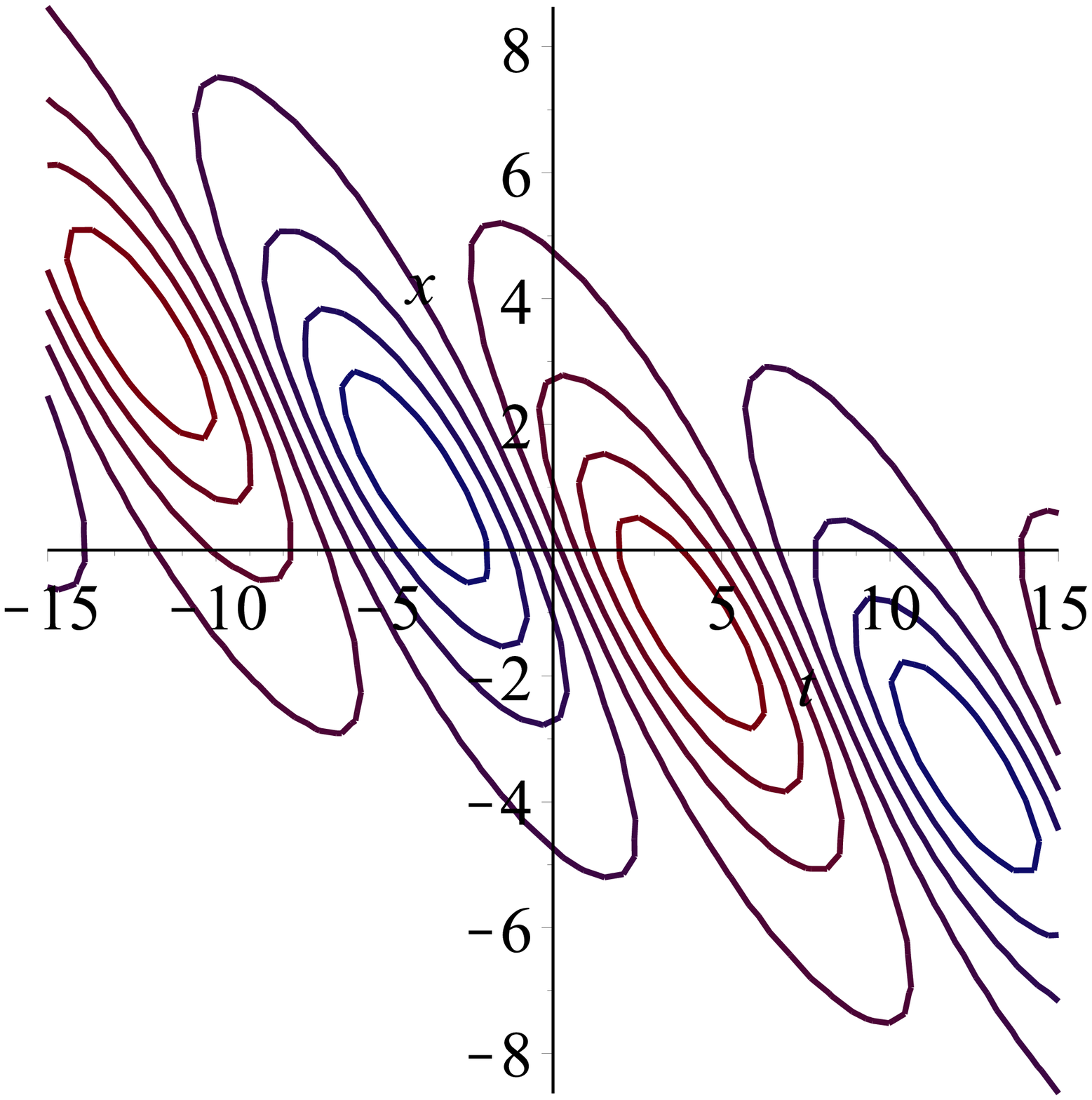}}}
\subfigure[]{\resizebox{0.28\hsize}{!}{\includegraphics*{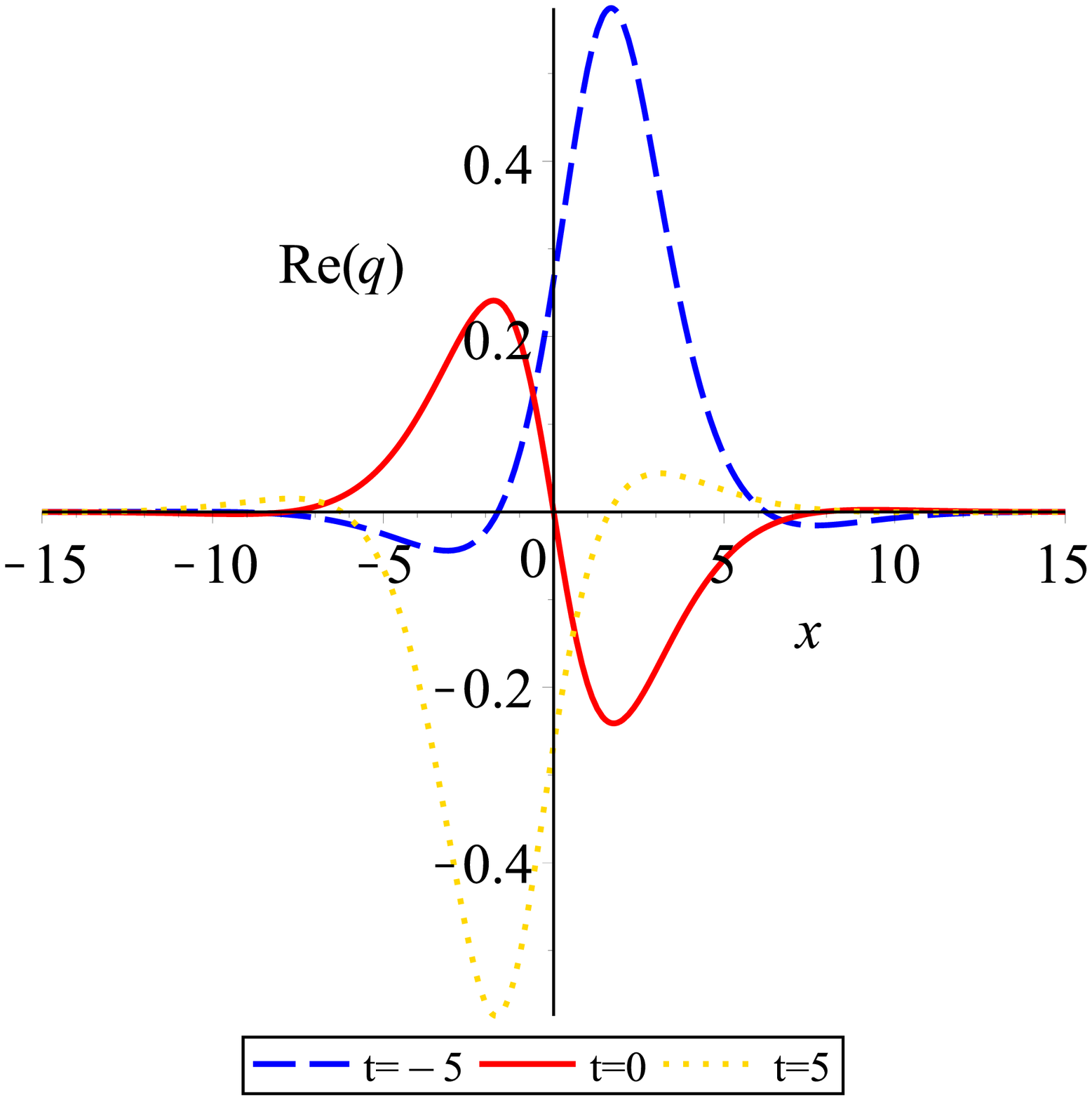}}}
\parbox[c]{13.0cm}{\footnotesize
{\bf Figure 2.}~Plots of one-soliton solution (26). (a) Perspective view of real part of $q$; (b) Contour plot of real part of $q$ in Figure 2(a); (c) The soliton along the $x$-axis with different time in Figure 2(a).}
\end{center}
\addtocounter{subfigure}{-3}
\end{figure}
\begin{figure}
\begin{center}
\subfigure[]{\resizebox{0.3\hsize}{!}{\includegraphics*{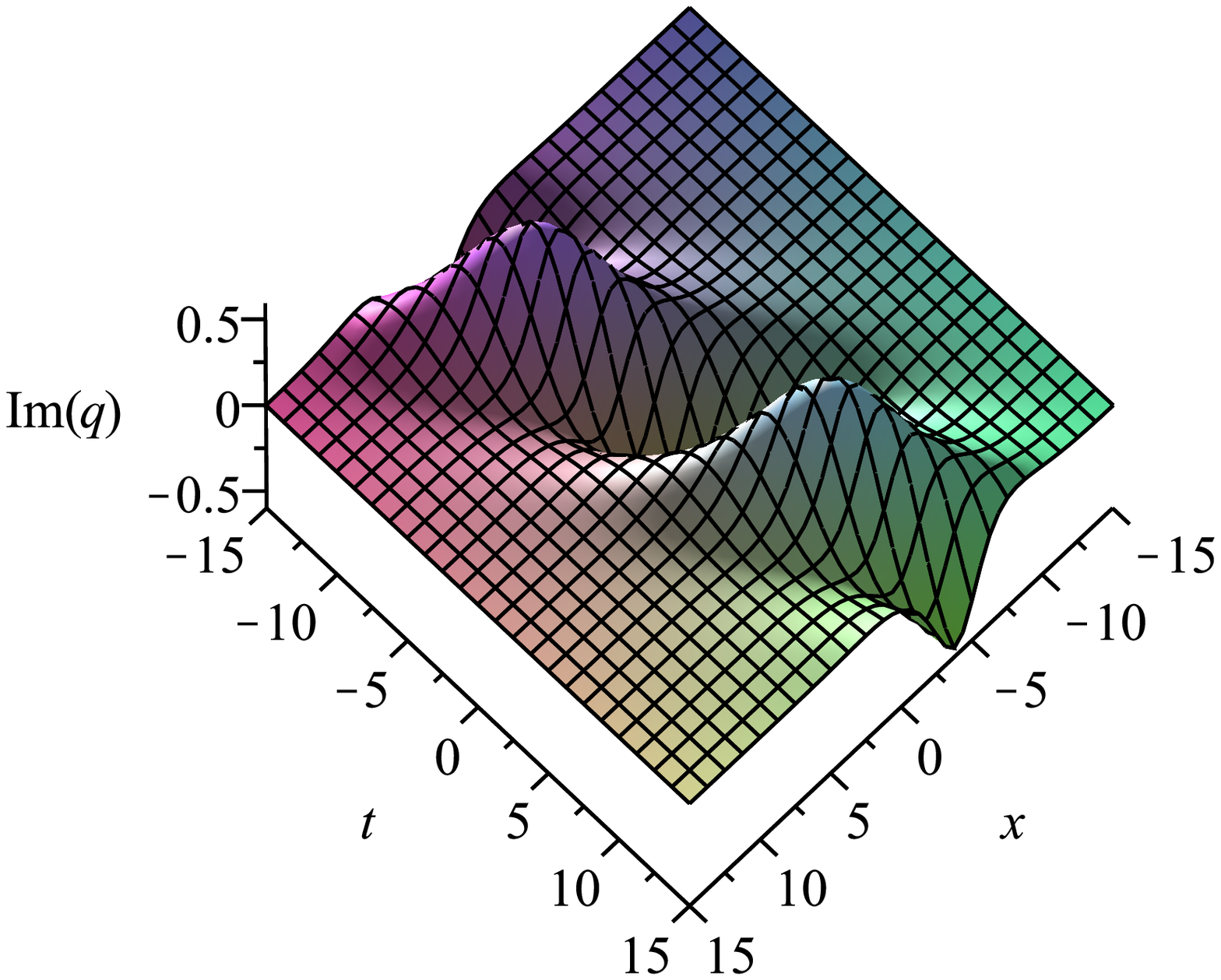}}}
\subfigure[]{\resizebox{0.28\hsize}{!}{\includegraphics*{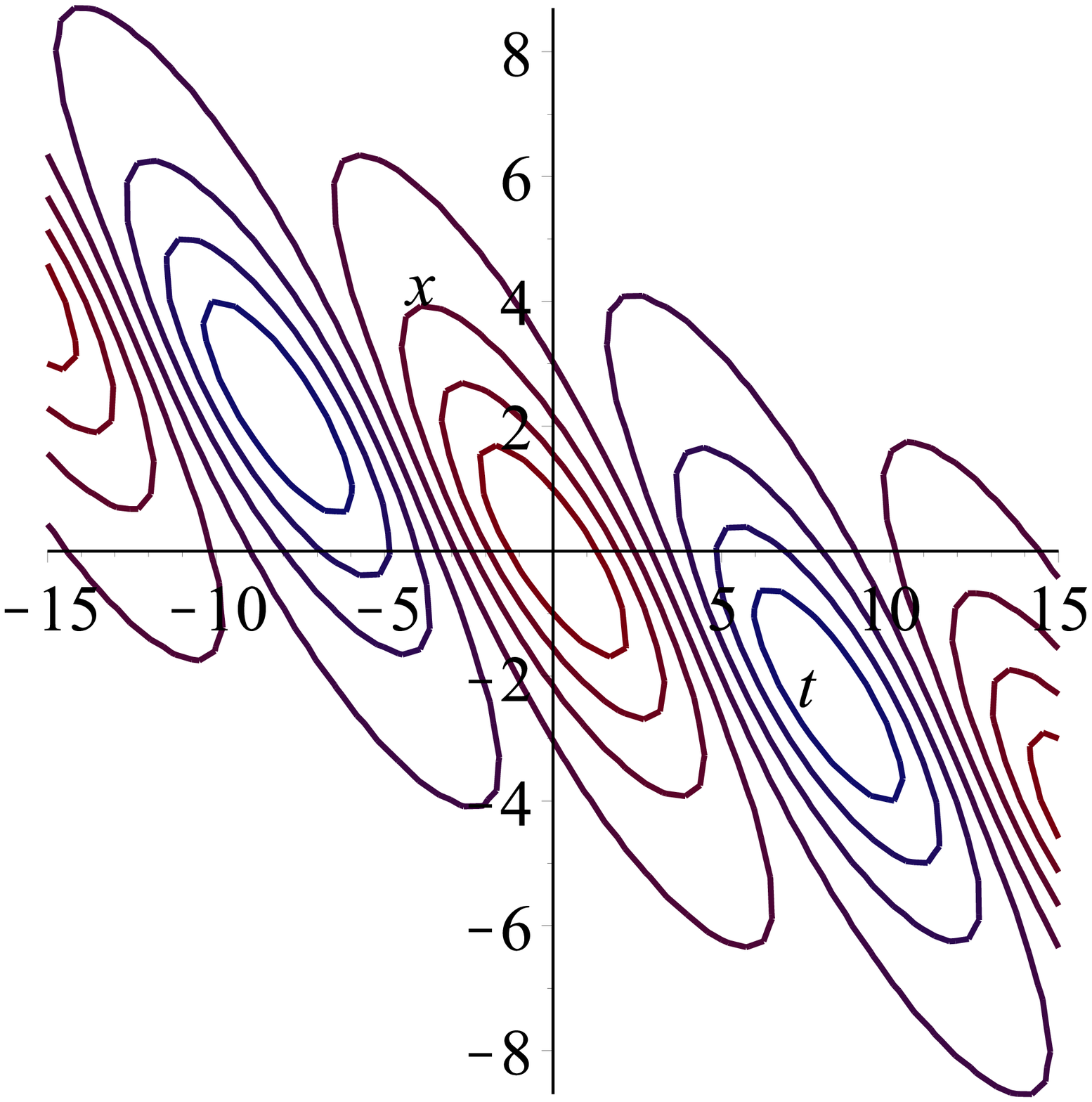}}}
\subfigure[]{\resizebox{0.28\hsize}{!}{\includegraphics*{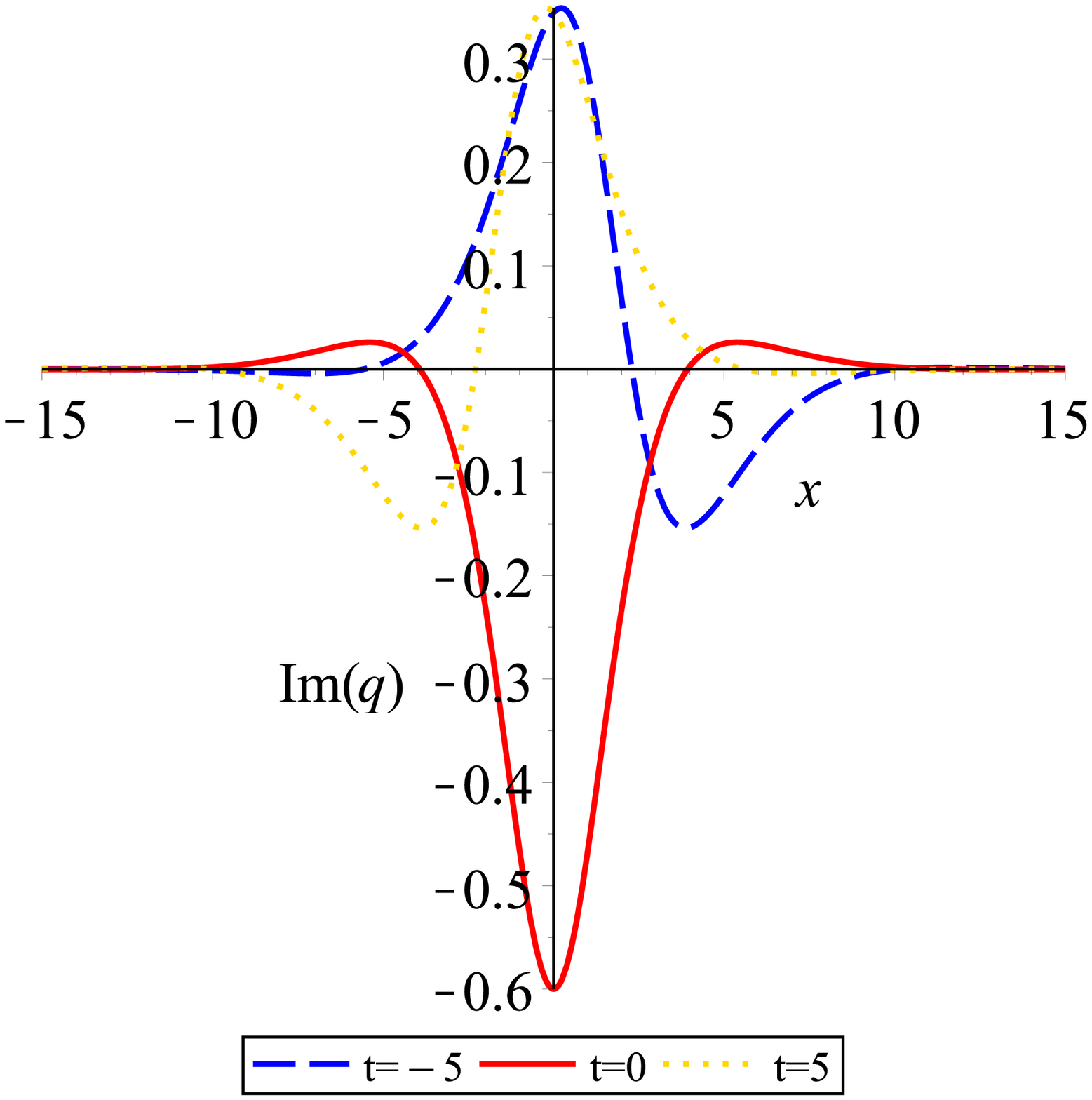}}}
\parbox[c]{13.0cm}{\footnotesize
{\bf Figure 3.}~Plots of one-soliton solution (26). (a) Perspective view of imaginary part of $q$; (b) Contour plot of imaginary part of $q$ in Figure 3(a); (c) The soliton along the $x$-axis with different time in Figure 3(a).}
\end{center}
\addtocounter{subfigure}{-3}
\end{figure}
\begin{figure}
\begin{center}
\subfigure[]{\resizebox{0.3\hsize}{!}{\includegraphics*{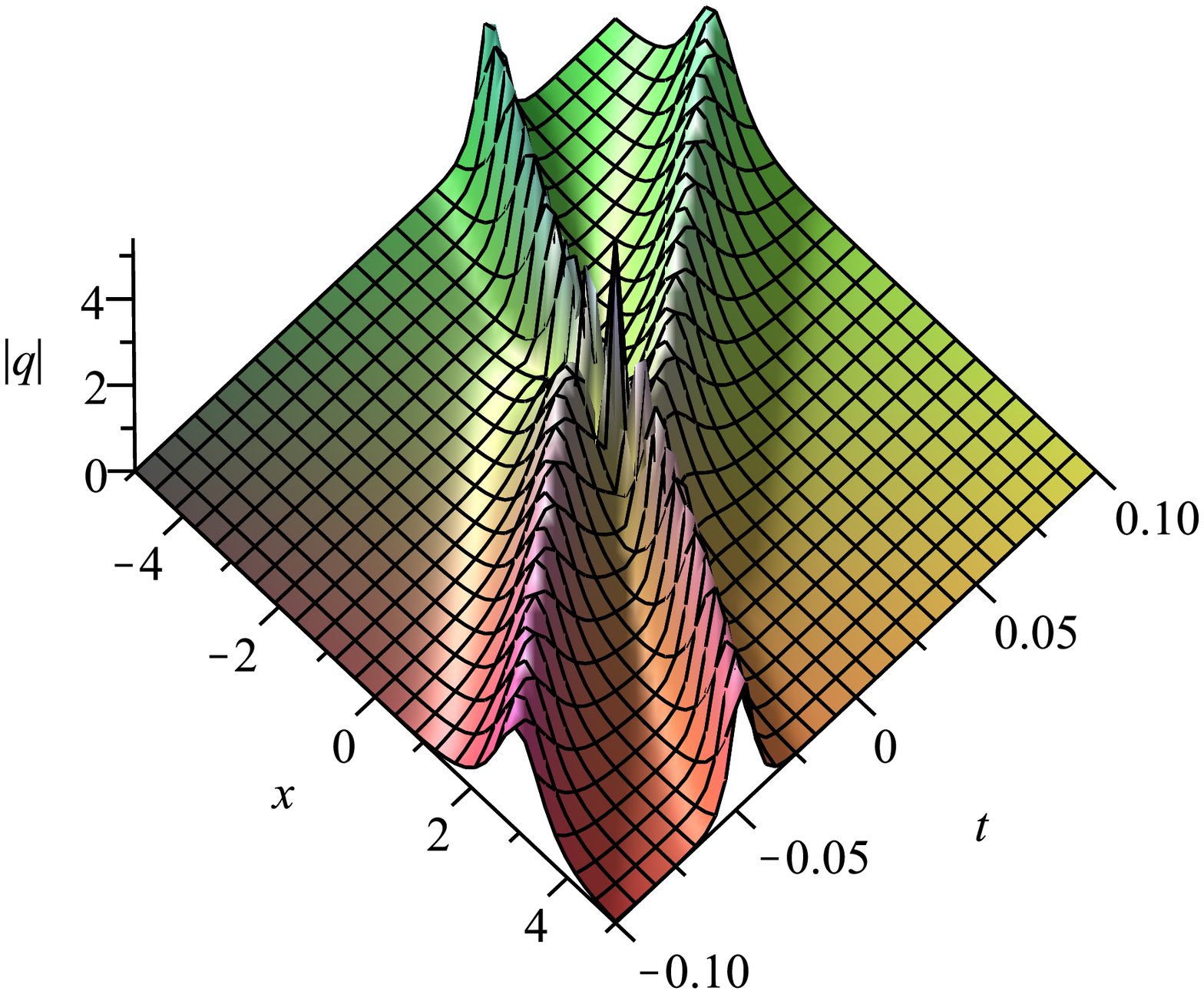}}}
\subfigure[]{\resizebox{0.32\hsize}{!}{\includegraphics*{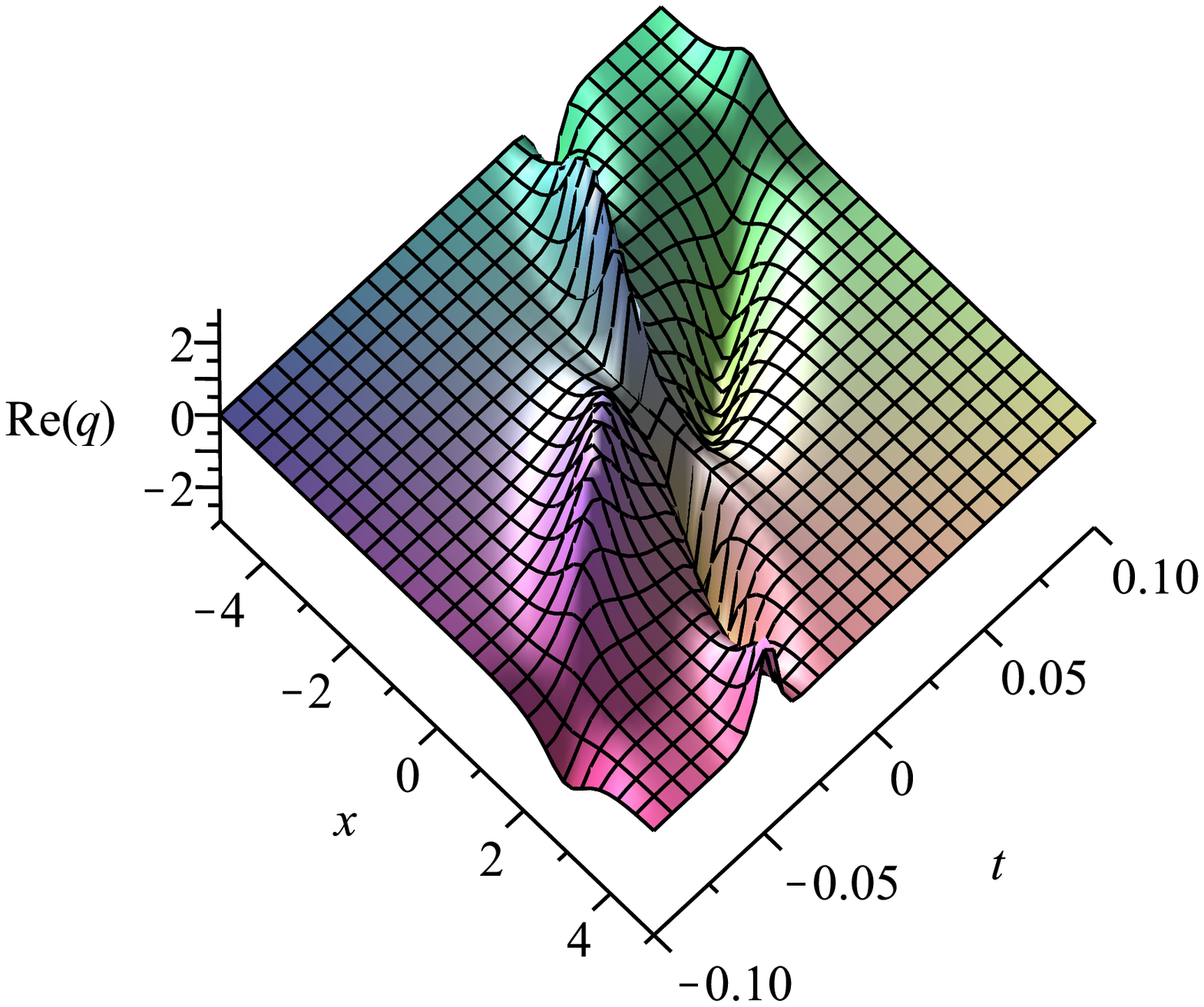}}}
\subfigure[]{\resizebox{0.32\hsize}{!}{\includegraphics*{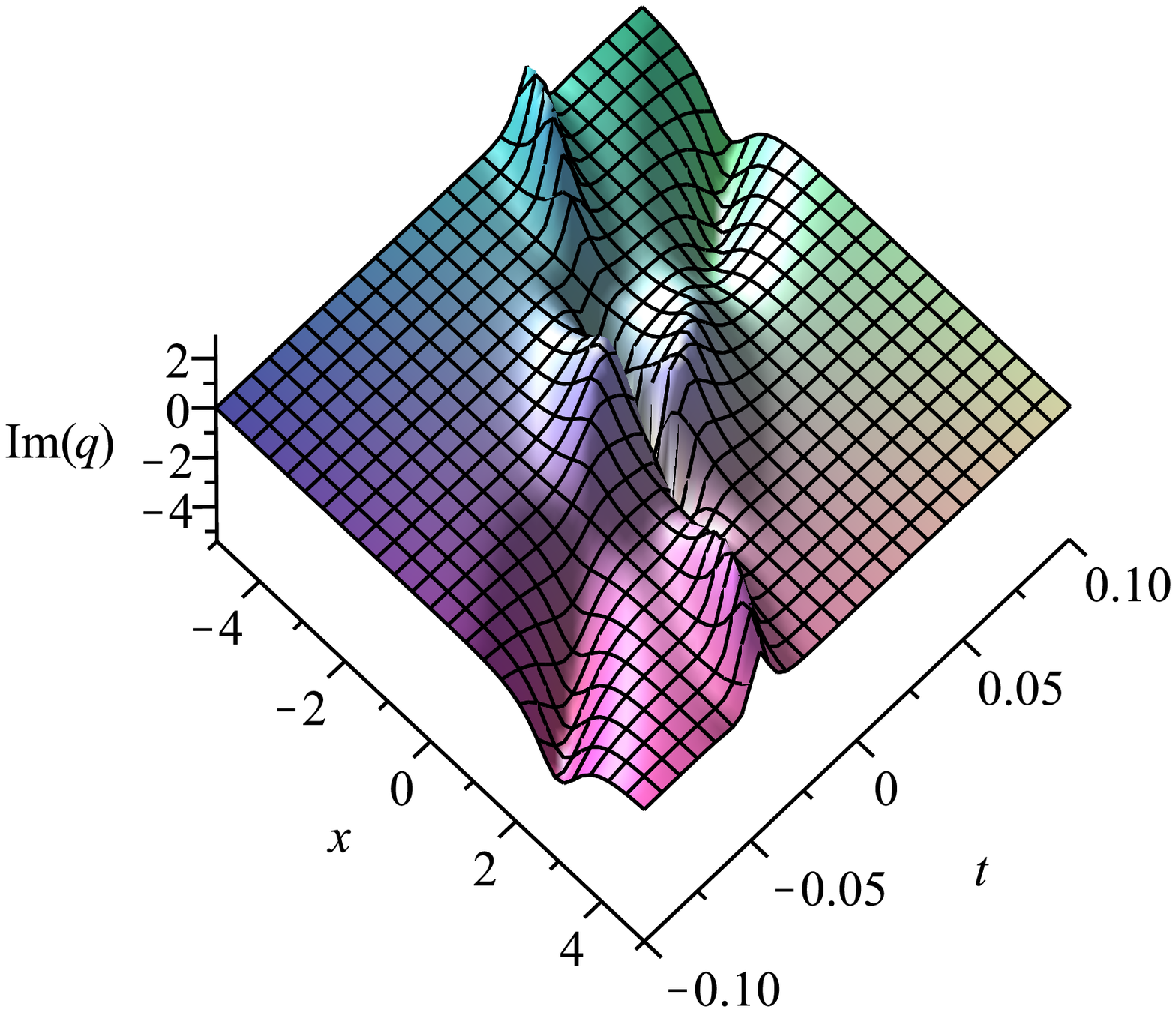}}}
\parbox[c]{13.0cm}{\footnotesize
{\bf Figure 4.}~Plots of two-soliton solution (28). (a) Perspective view of modulus of $q$; (b) Perspective view of real part of $q$; (c) Perspective view of imaginary part of $q$.}
\end{center}
\addtocounter{subfigure}{-3}
\end{figure}
\begin{figure}
\begin{center}
\subfigure[]{\resizebox{0.3\hsize}{!}{\includegraphics*{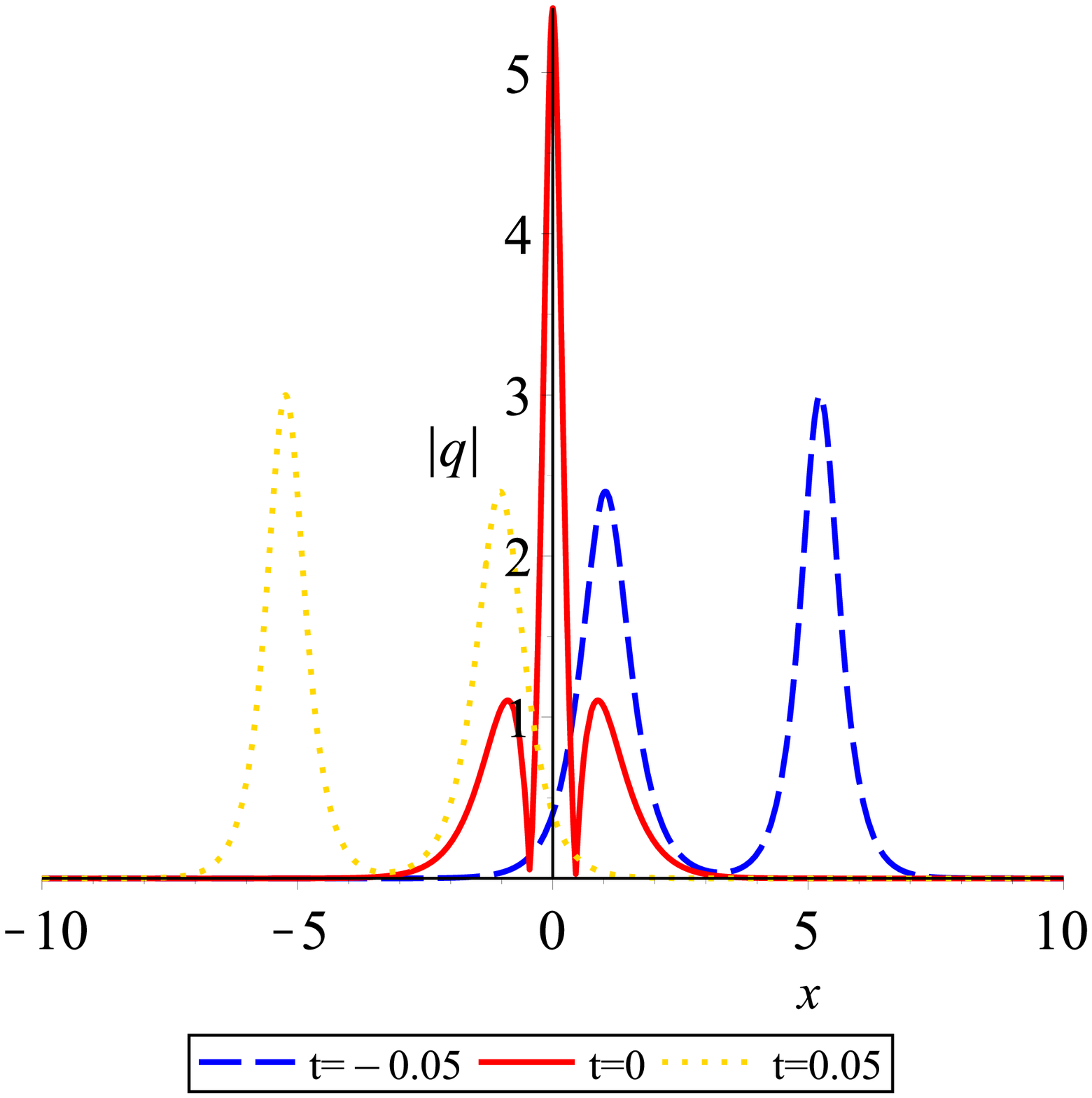}}}
\subfigure[]{\resizebox{0.3\hsize}{!}{\includegraphics*{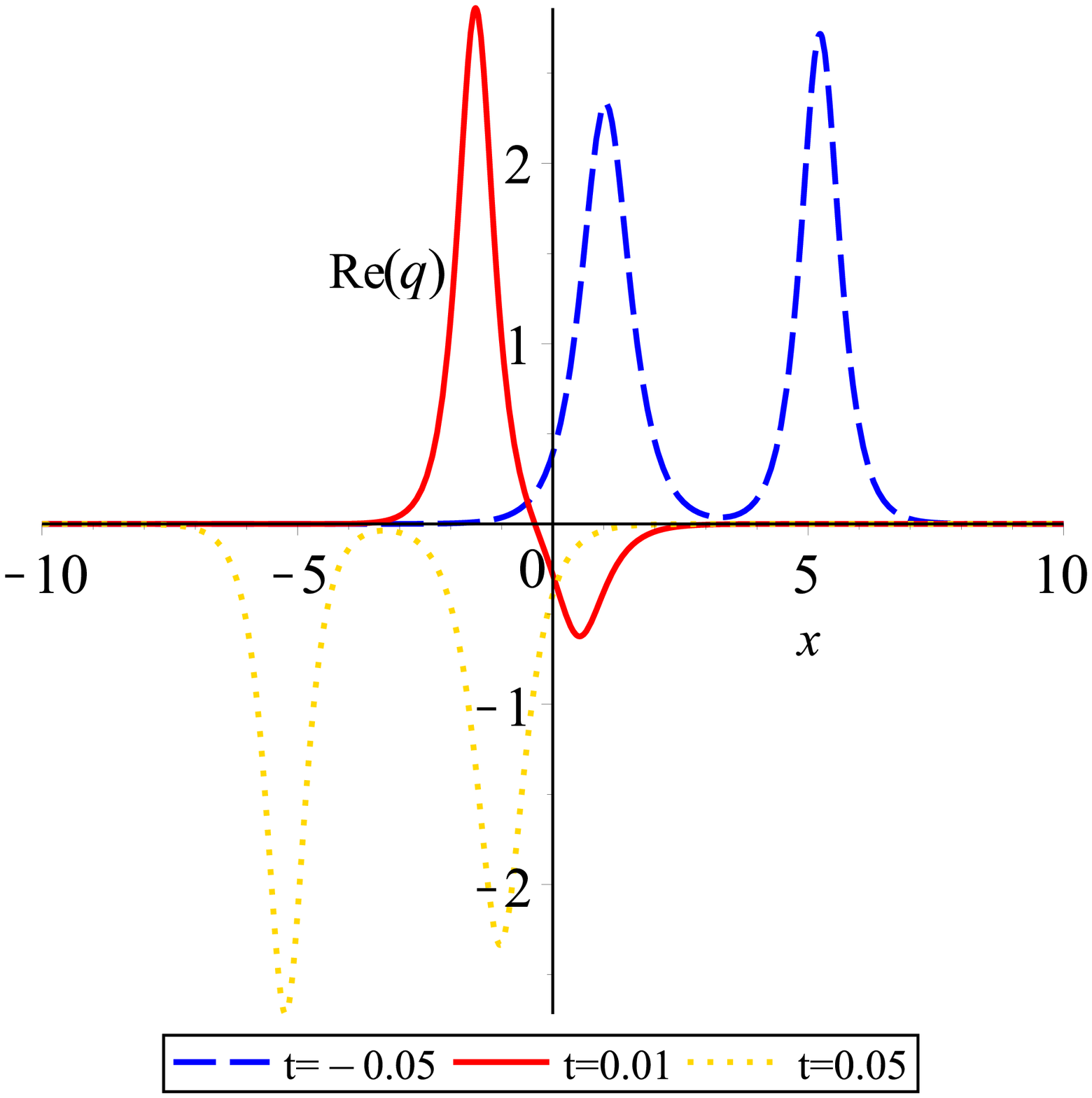}}}
\subfigure[]{\resizebox{0.3\hsize}{!}{\includegraphics*{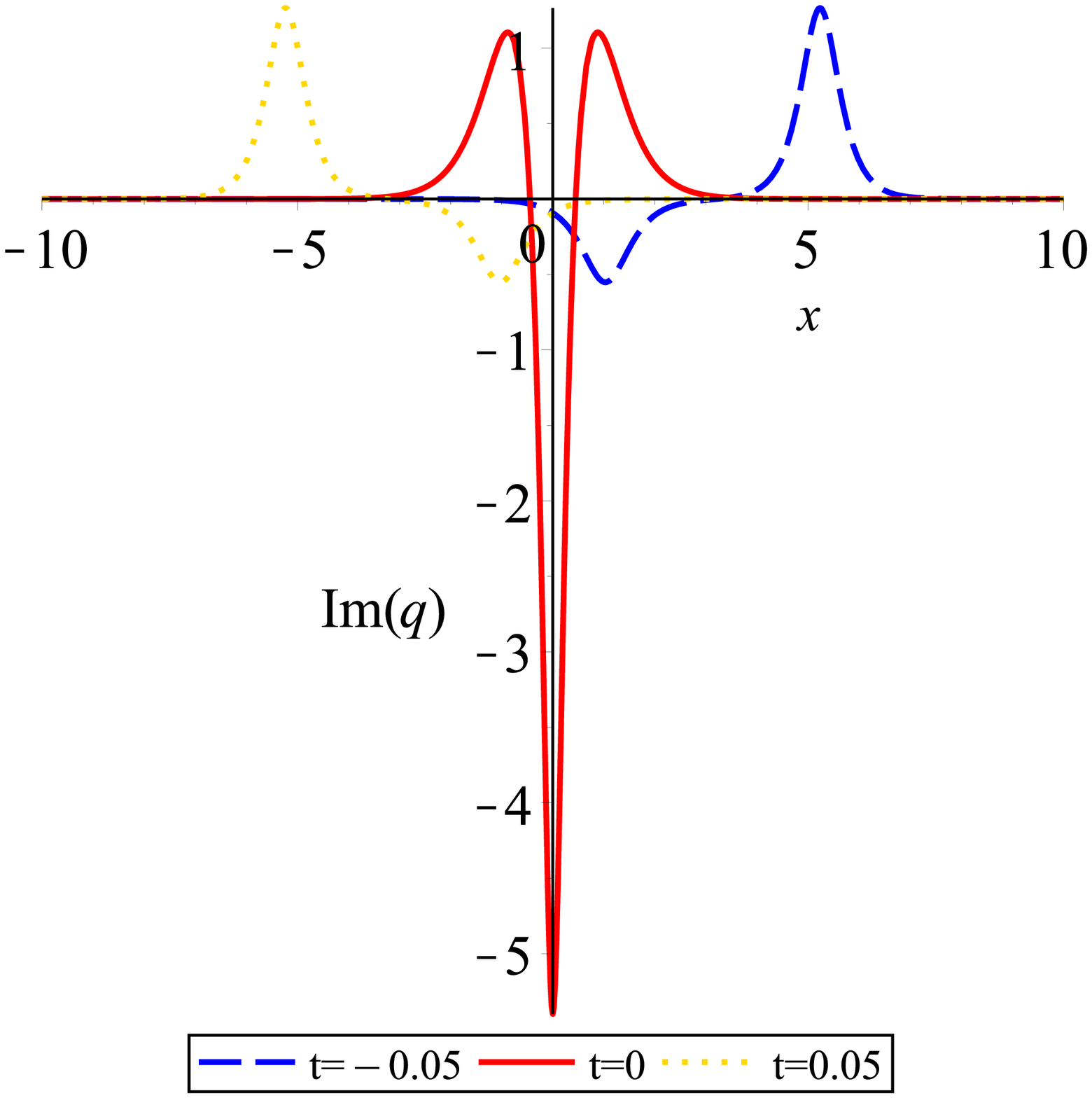}}}
\parbox[c]{13.0cm}{\footnotesize
{\bf Figure 5.}~Plots of two-soliton solution (28). (a) The soliton along the $x$-axis with different time in Figure 4(a). (b) The soliton along the $x$-axis with different time in Figure 4(b); (c) The soliton along the $x$-axis with different time in Figure 4(c).}
\end{center}
\end{figure}

\section{Conclusion}
The aim of the paper was to investigate a fifth-order nonlinear Schr\"{o}dinger equation describing the one-dimensional anisotropic Heisenberg ferromagnetic spin chain via the Riemann-Hilbert approach. The spectral analysis of the associated Lax pair was first carried out and a Riemann-Hilbert problem was established. After that, via solving the presented Riemann-Hilbert problem with reflectionless case, the $N$-soliton solution to the fifth-order nonlinear Schr\"{o}dinger equation were attained at last. Furthermore, by selecting specific values of the involved parameters, a few plots of one- and two-soliton solutions were made to display the localized structures and dynamic behaviors.





\end{document}